\documentclass[journal]{IEEEtran}
\usepackage{cite}
\usepackage{longtable}
\usepackage{xcolor,soul,framed} 

\colorlet{shadecolor}{yellow}
\usepackage{graphicx}
\usepackage{epstopdf}
\graphicspath{{../pdf/}{../jpeg/}}
\DeclareGraphicsExtensions{.pdf,.jpeg,.png}

\usepackage{amsmath}
\usepackage{enumitem}
\usepackage{array}
\usepackage{amsmath}
\usepackage[english]{babel}
\usepackage[utf8]{inputenc}
 
 \usepackage{float}
\usepackage{xcolor}
\usepackage{fancyhdr,graphicx,amsmath,amssymb}

\usepackage{xcolor,soul,framed} 
\colorlet{shadecolor}{yellow}

\usepackage{array}
\usepackage{mdwmath}
\usepackage{mdwtab}
\usepackage{eqparbox}
\usepackage{url}
\usepackage{soul}
\usepackage{algorithm}
\usepackage{algorithmic}
\makeatletter

\usepackage{amsmath}

\ifCLASSINFOpdf
\else
\fi

\begin{document}

\title{Toward Experience-Driven Traffic Management and Orchestration in Digital-Twin-Enabled 6G Networks}

		\author {Muhammad Tariq,~\IEEEmembership{Senior Member, ~IEEE}, Faisal Naeem,~\IEEEmembership{Member,~IEEE,} and H. Vincent Poor,~\IEEEmembership{Life Fellow,~IEEE}
		\thanks{Muhammad Tariq si with the Department of Electrical Engineering, National University of Computer \& Emerging Sciences, Islamabad, Pakistan. E-mail: tariq.khan@nu.edu.pk}
		\thanks{Faisal Naeem is with the Department of Electrical Engineering, University of Quebec, Montreal, Canada. Email: faisal.naeem.1@ens.etsmtl.ca}
		\thanks{H. Vincent Poor is with the Department of Electrical and Computer Engineering, Princeton University, 08544, Princeton, NJ, USA, Emails: poor@princeton.edu}
		\thanks{\emph{Corresponding author is Muhammad Tariq}}
	}

\maketitle

\begin{abstract}
 The envisioned 6G networks are expected to support extremely high data rates, low-latency, and radically new applications empowered by machine learning. The futuristic 6G networks require a novel framework that can be used to operate, manage, and optimize its underlying services such as ultra-reliable and low-latency communication, and Internet of everything. In recent years, artificial intelligence (AI) has demonstrated significant success in optimizing and designing networks. The AI-enabled traffic orchestration can dynamically organize different network architectures and slices to provide quality of experience considering the dynamic nature of the wireless communication network.
 In this paper, we propose a digital twin enabled network framework, empowered by AI to cater the variability and complexity of envisioned 6G networks, to provide smart resource management and intelligent service provisioning. Digital twin paves a way for achieving optimizing 6G services by creating a virtual representation of the 6G network along with its associated communication technologies (e.g., intelligent reflecting surfaces, terahertz and millimeter communication), computing systems (e.g., cloud computing and fog computing) with its associated algorithms (e.g., optimization and machine learning). We then discuss and review the existing AI-enabled traffic management and orchestration techniques and highlight future research directions and potential solutions in 6G networks.
\end{abstract}

\begin{IEEEkeywords}
Software Defined Networking, Deep Reinforcement Learning, 6G, Ultra-reliable and Low-latency Communication, Digital Twin
\end{IEEEkeywords}

%
\IEEEpeerreviewmaketitle
\section{Introduction}

The dramatic development of adaptive and complex communication technologies has triggered the emergence of new architectures such as Internet of Things (IoT), Intelligent transportation systems (ITS), smart healthcare, and Intelligent reflecting surfaces (IRSs). Specifically, the next-generation networks incorporating the IRS-enabled systems are based on radical new applications that have their performance requirements, such as Internet of everything (IoE), ultra-reliable and low-latency communication (uRLLC), massive machine-type communication (mMTC), and  enhanced mobile broadband (eMBB). The 5G wireless networks have not been able to fully address the challenges and requirements, brought by the massive data and real-time requirements of services \cite{tariq2020vulnerability}.

In contrast to the previous generations, envisioned 6G networks promise to revolutionize and transform the wireless networks from \textit{``connecting things"} to \textit{``connected intelligence"}. Even though the development of 6G network is at the initial stage, however, it intends to achieve very high data rates, up to 1 Tb/s,  end-to-end delay requirements (less than 1 msec), and connected intelligence with artificial intelligence (AI) capability. Moreover, the network's capacity requirement is much higher than the 5G networks and has to utilize the frequency spectrum in the range of 100 GHz to 3 GHz. Nevertheless, besides the network capacity and end-to-end delay requirements,  IRS-enabled unmanned aerial vehicle (UAV) and terrestrial networks will have capacity and energy-efficiency requirements. As a result more detailed requirements are needed in the context of traffic optimization considering the dynamic and complex environment of the envisioned 6G networks.

Network traffic control (NTC) can play a vital role in designing routing optimization techniques with diverse network requirements and varying channel conditions in 6G networks. However, the \textit{Best-effort} architecture cannot provide flexible network management in highly mobile and dynamic networks. The \textit{programmability} feature of the software-defined networking (SDN) provides intelligence in the network and can provide better incentives for NTC to optimize the traffic in 6G networks. NTC techniques are usually formulated as an optimization problem and cannot accurately model the dynamic and time-varying channel conditions for the anticipated 6G networks. Thus, intelligent and adaptive techniques are required that can satisfy the network requirements of the envisioned 6G networks.

\begin{figure*}
  \begin{center}
  \renewcommand\figurename{Fig.}
    \includegraphics[width=5.5 in, height= 3.5 in]{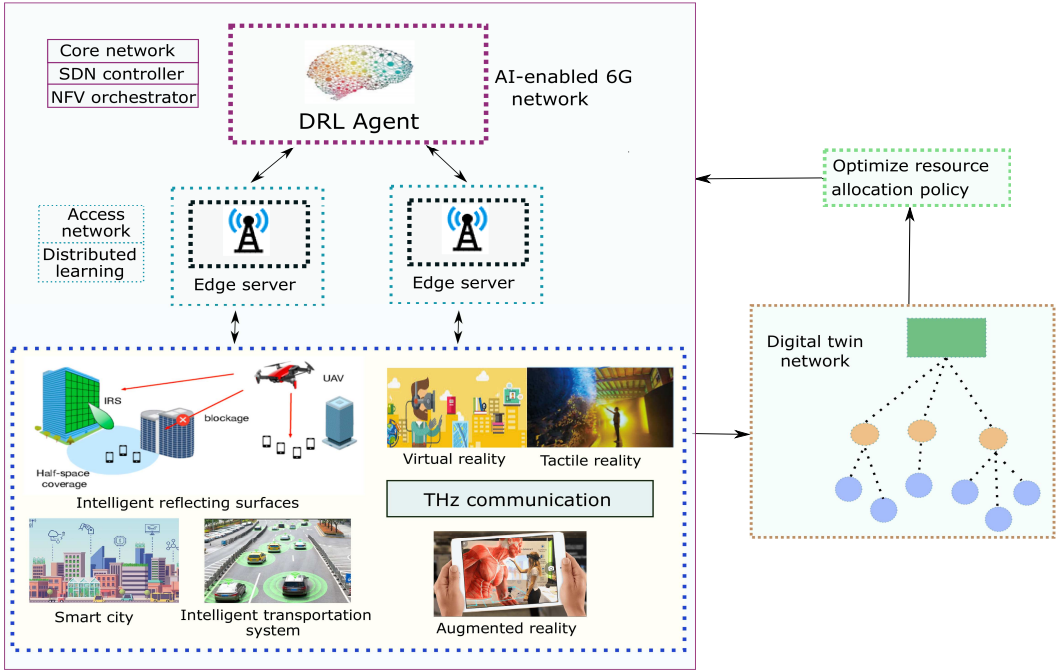}\\
     \caption{RL-aided architecture for traffic optimization in 6G Networks}
   \end{center}
\end{figure*}

Machine learning (ML) is regarded as a paradigm shift in achieving intelligence in the networks for the envisioned 6G networks. Reinforcement learning (RL) is an AI-based algorithm that can achieve human-level intelligence. In RL, a smart agent interacts with the environment in an episodic manner, and as a result, it receives the reward and next state to learn the optimal policies. In the context of 6G networks, it is expected that they will need to operate in highly mobile and complex environments especially in the case where IRSs will be deployed in a terrestrial and UAV-enabled networks. TRL can enable the next-generation networks such as those incorporating IRSs to intelligently and actively monitor their environments by learning the optimal policies and proactively taking the actions to maximize a pre-defined reward. Moreover, RL-aided decision-making is a key-enabler for 6G services, particularly those that have stringent QoS requirements and highly dynamic wireless propagation environments. In fact, if well designed, RL optimization techniques provide self-healing, self-optimizing, and self-organizing solutions for a broad range of applications of 6G within the context of traffic optimization.

A digital twin (DT) is a technology that creates the virtual representation of the dynamics and elements of a physical system. DT uses data analytic and ML techniques to learn the dynamic characteristics of a communication network. DT can play a vital role in traffic optimization by creating a virtual representation of a physical network depending on implementation factors (such as IRSs, multiple-input-multiple-output MIMO, UAVs etc.,). It captures the time-varying characteristics between the communication network and surrounding network. The combination of SDN, DT, network function virtualization (NFV), and network slicing (NS) with RL can achieve zero-touch network orchestration, management, and optimization, which can be an effective strategy for the paradigm shift from 5G to 6G networks. RL-aided network orchestration can dynamically orchestrate network slices and architecture and satisfy the dynamic applications and services. RL-aided network optimization can globally monitor a real-time network's key performance indicators (KPI) and adaptively learn the network policies to satisfy the KPIs continuously. As a result, the RL and DT will play a significant role in learning the optimal routing policies in 6G networks. 

The key contribution of the paper are as follow.
\begin{enumerate}
\item Realizing the network intelligence capabilities in envisioned 6G networks for traffic optimization, an AI-assisted intelligent digital-twin enabled traffic orchestrator framework for traffic management in proposed.
\item Moreover, a review of the recent AI-enabled traffic orchestration techniques in envisioned 6G networks, and the challenges and future research directions are discussed in detail at the end.
\end{enumerate}

\section{PROPOSED ARCHITECTURE}
In this section, we proposed the most eye-catching digital twin-empowered architecture for traffic optimization in envisioned 6G networks as shown in Fig.1. The proposed RL-enabled and digital twin empowered architecture is mainly divided into four layers: smart application layer, digital twin layer. data analytic layer, and intelligent control layer.
\begin{enumerate}
    \item \textit{Smart application layer}: This layer is responsible for delivering the application-dependent requirements of smart applications such as IRSs, smart cities, ITS, virtual and tactile reality. The real-time networks in 6G such as, extremely reliable and low-latency communications (ERLLC), extremely low-power communications (ELPC), eMBB and mMTC, will forward their key performance indicators (KPI) using the smart application layer.

     \item \textit{Digital twin layer}:  The digital twin layer enables digital representation of the physical systems and virtual model of the physical models. The proposed digital twin layer will mirror the dynamic characteristics of the 6G-enabled physical wireless communication network based on the type of implementation (IRS, MIMO, UAV etc.,) and will make predictions. The digital twin in the proposed architecture will construct the dynamic characteristics of the physical communication network, training of machine learning parameters, i.e., stochastic task arrival process and dynamic changes in network resources. Finally, the digital twin objects will be created dynamically to optimize resource allocation and offloading policy of the 6G applications such as mMTC, eMBB, and ERLLC.

    \item \textit{Distributed learning}: The network control system has gone through a centralized approach to the distributed structure. The envisioned 6G network will have diverse network requirements such as uRLLC where the single agent controller cannot optimize the resources of the network efficiently. Furthermore, to increase the computing resources and intelligence in the network, advanced distributed learning techniques will be integrated in the smart agents such  multi-access edge computing (MEC) with RL-aided algorithms that can be utilized for optimizing the routing for smart applications (e.g., transportation, agriculture and healthcare).
    
    \item \textit{Intelligent control layer}: The combination of AI and SDN/NFV/NS in the intelligent control layer can achieve optimization, orchestration and a flexible decision making process in envisioned 6G networks. The smart agents in the distributed layers forward the learned optimal policy to the intelligent control plane to intelligently learn and optimize the global optimal policies. The multi-agent reinforcement learning (MARL) learns the policy in a network based on joint actions of all the agents. In the MARL process, the training is done in the centralized approach and execution of the policy is done in a distributed approach. Thus, the SDN controller will play a vital role in centralized training of MARL algorithms for the traffic optimization.\
    \item \textit{Data-plane layer}:
    The control plane learns the optimal traffic routing policies while the data-plane layer acts as a forwarding device and is used for forwarding the traffic to the end-users.

\end{enumerate}

\section{Recent Advances In RL-aided Traffic Optimization}

More recently, the model-free advanced ML algorithms such as RL and DRL have been incorporated in the control plane for traffic optimization. The RL approach learns the optimal routing optimization policy from its observation by interacting with the network environment. These approaches are not dependent on pre-defined policies and have the ability to intelligently adapt to the dynamic environment. We provide a comprehensive literature review of the exiting RL-aided techniques used for traffic optimization using the SDN approach and discuss whether they can be applied to the envisioned 6G networks or not.

\subsection{Single-agent reinforcement learning for traffic optimization}
The single agent RL algorithms perform well for a small-network. However, the performance degrades when the scale of the network increases. A single agent deep reinforcement learning (DRL) is proposed for solving the traffic optimization issue in case of wireless communication networks. However, the next-generation networks such as 6G will have complex, continuous state and action variables, where DRL approach cannot perform well.
In \cite{huang2018deep}, the authors used a DRL based deep deterministic policy gradient (DDPG) algorithm for optimizing the routing in SDN scenario. The actor-critic based DDPG framework is used for solving the quality of service (QoS) and QoE in networks. The AC-based DDPG algorithm performs well compared to the DRL technique.\\
The authors proposed a DDPG model \cite{liu2020drl}, which uses a convolutional neural network (CNN) as a function approximator for improving the routing performance. The paper combines multiple network resources such as (bandwidth and cache) by improving their contribution in reducing the delay.
The traffic engineering (TE) problem using the DDPG algorithm cannot have satisfactory performance due to its uniform sampling technique for experience replay, which ignores the importance of transition samples in the replay buffer. Thus, DDPG framework cannot perform well for the traffic optimization problems.
In \cite{sun2019tide}, the authors proposed a recurrent neural network-based DRL technique for routing optimization in an SDN-enabled network. The algorithm is implemented on a test-bed and simulation result shows that the proposed long short-term memory (LSTM) based DRL reduces the traffic delay and loss as compared to the traditional open shortest path first (OSPF) routing technique. This approach does not perform better when an agent is introduced to a new environment. The TE problem based on DDPG are used for global offline routing as the greedy online QoS optimization poses issues such as avoiding network loops, variance in selecting optimal actions and learning the optimal paths considering multiple QoS metrics. To solve the issue, the intelligent nature of the SDN controller implements and calculates the routing policies on a real-time and forwards the policies to the switches. The decoupled nature of the control plane provides flexibility and data management and performs several functionalities such as traffic detection, traffic prediction and resource management. The envisioned 6G networks will generate high dimensional states from heterogeneous networks, where the existing single controller-based RL algorithms will not be able to manage large number of flows arriving at a time on controller and can result in a switch migration problem (SMP). Thus, a multi-agent reinforcement learning based traffic optimization techniques are needed.

\begin{table*}[ht]
\centering
  \caption{Comparison of different Reinforcement Learning algorithms for traffic optimization using SDN approach}
  \renewcommand{\arraystretch}{2}
  \begin{tabular}{ |p{1.5cm}|p{6 cm}|p{5cm}|p{5cm}|}
  \hline
  References &	Key features &	 Pro   &	Cons \\
  \hline
  \hline
 
  \cite{huang2018deep} &	Utilize the actor and critic networks to train the network &	Better storage overhead and improved convergence time &		Control strategies and parameters not well-defined \\
   \hline
   
   \cite{liu2020drl} & DDPG with CNN function approximator & Achieves low flow completion time &  Collecting fine grained network statistics.\\

 \hline

   \cite{sun2019tide} & An intelligent network control architecture for optimizing QoS & Tested on real test-bed and improved network transmission delay, & Communication overhead and performance degrades when the topology changes.\\
   \hline
   \cite{jalil2020dqr} & A dueling deep Q-network based on prioritised experience replay for routing optimization & Optimise multiple QoS constraints and avoids network loops & Requiring the information of important experience is difficult.\\
   \hline
   \cite{chen2020rl} &  A dueling DDQN for achieving optimizing QoS and achieving scalability & Network can optimize upward and downward traffic characteristics &  Less efficient with small state-action pairs.\\

   \hline
    \cite{sun2020scalable} & Actor-critic with pinning control theory to reduce the large dimensionality issue & Achieves less end-to-end delay and adaptive to large networks & Control theory with DRL not explored for optimal performance.\\
    \hline

    \cite{sun2020marvel} & A distributed multi-agent approach for solving the switch mitigation problem (SMP) & Improved processing time and reduced control plane requests. & Very high computational cost.\\
    \hline
    
    \cite{geng2020multi} & Designed a multi-agent reinforcement learning technique for reducing congestion in multiple regions & Achieved scalability and reduced congestion & Performance degrades when network topology and the QoS parameters changes.\\
    \hline

    \cite{pinyoanuntapong2019distributed} & A multi-agent based actor-critic approach for multi-path routing &  Minimize E2E delay and efficiently load balancing &  Not tested on large-scale network. \\
    \hline

    \cite{zhao2020improving} & A multi-agent RL framework for improving the inter-domain routing & Scalable and achieves higher throughput & Cannot learn the non-linear information between inter-domain networks.\\
    \hline


     \hline
     \cite{lei2020congestion} & A multitask DRL approach for traffic management & It can optimize congestion and load-balance simultaneously. & Training process is slow and a transparent DRL congestion model is required.\\
     \hline
    
     \cite{naeem2020generative} & A GAN based deep distributional Q learning for transmission scheduling &  Higher spectral efficiency and received service level agreement under random noise & Not tested under multi-constrained QoS parameters.\\
     \hline

  \end{tabular}
\end{table*}

\subsection{Multi-agent reinforcement learning for traffic optimization}
The authors in \cite{sun2020marvel} proposed a distributed control plane as a multi-agent approach and formulate the multi-agent reinforcement learning (MARL) to solve the scalability and load balancing issue in large-scale networks. The MARL is used for learning the optimal policies in distributed systems. In the distributed control plane, each control plane acts as a DRL agent and all agents in the network forms a multi-agent system. Each agent in the network interacts with its neighbour and exchange the information with other agents. Finally, the important information is retrieved from all agents and a global optimal policy is learnt for flexible load management. The paper also designed a zero-sum game technique for the MARL for the stability of the model. The results show that MARL reduced the processing time of controller by 25\% and improves the overhead ratio generated through the control plane request by 27.3\%.

The existing RL techniques for TE for multi-region network learn the optimal policy based on its own regional network states and observations. To jointly learn the global behavior for TE, it may result in scalability and overhead issues. Thus, designing the distributed TE in multiple heterogeneous networks is very challenging issue in contrast to the traditional single network. To solve the overhead issue, the authors in \cite{geng2020multi} proposed a multi-agent based DRL for solving the TE problems in multiple heterogeneous networks. The framework is based on two reinforcement learning agents known as T-agents and O-agents implemented for each region. The T-agent learns the optimal network policy within a local network to forward the terminal traffic for satisfying a local TE requirements. The second agent learns the routing policy from neighbouring regions to forward the outgoing traffic for optimizing a cooperative and global TE objective. The simulation result shows that multi-region RL techniques can achieve optimal performance and reduce the congestion with achieving robustness and scalability.

\begin{table*}[ht]
\centering
  \caption{Summary of DRL model and algorithms for traffic optimization}
  \renewcommand{\arraystretch}{2}
  \begin{tabular}{ |p{1cm}|p{2 cm}|p{5cm}|p{5cm}|p{3.5cm}|}
  \hline
  References &	DRL Model &	 State & Action & Rewards  	 \\
  \hline
   \hline
   \cite{liu2020drl} & DDPG with CNN & Resource allocation and resource demand state & Select end-to-end paths & Throughput and flow completion time and throughput \\
   \hline
  
  \hline
  \cite{sun2019tide} & LSTM-DDPG & Delay, jitter, throughput and loss & Adjust link weight for each source and destination pair & QoS feedback from network \\
  \hline
  \cite{jalil2020dqr} & DDQN &   Loss, delay and bandwidth & Tunable weights to prioritize QoS metric & Penalty for selecting wrong actions and vice versa \\
  \hline
  \cite{chen2020rl} & Dueling DDQN & link throughput, link delay, link capacity, link trust level and switch statistics & Choose the path that minimizes upward and downward delay & Throughput rate and delay rate \\

  \hline
   \cite{sun2020scalable} & Actor-critic with pinning control theory & Traffic distribution on each link & Select k-critical flows & Average end-to-end delay and maximum link utilization ratio \\ 
   \hline
   
   \cite{sun2020marvel} & LSTM based DRL for Multi-agent reinforcement learning problem & Resource utilization of controllers in the network & Minimum and maximum resource utilization & Improving the resource utilization and load balancing \\ 
   \hline

   \cite{geng2020multi} & DDPG based multi-agent reinforcement learning & Inward and outward edge utilization of the region & Splitting ration of the traffic & To minimize edge cost of the network \\
   \hline

   \cite{pinyoanuntapong2019distributed} & Multi-agent Actor-critic & Source and destination, queue size, status of links and topology of network. & Selecting next-hop router & End-to-End delay \\
   \hline

   \cite{zhao2020improving} & Advantage actor-critic & End-to-end flow of Autonomous systems & Choosing next-hop & Maximize the average throughput of all flows. \\
   \hline


  \hline
   \cite{lei2020congestion} & DDPG with CNN layers & Congestion control and load-balancing state & For congestion control action is sending rate or nodes and for load balancing action is optimal path selection & For CC queue length and, bandwidth, and for load balancing total throughput and latency \\
   \hline
    
    \cite{naeem2020generative} &  A distributional deep Q network based on GAN & Status of the channel (idle or busy),
 priority level of channel access, quality of the channel, traffic
load on the channel, service level agreement, number of packets
arrived on a channel  &  Power consumption control
(active or sleep), spectrum management
( hand-off or wait), selection of
transmission modulation (adaptive modulation or other
schemes), and bandwidth allocated to
each packet &  Maximize the system utility\\
\hline

  \end{tabular}
\end{table*}

In\cite{pinyoanuntapong2019distributed}, the authors presented a distributed multi-agent based TE solution to learn the optimal routing policy for the multi-path routing. The authors formulated the multi-path routing problem as a multi-agent Markov decision problem and proposed a multi-agent actor-critic technique where each router functions as a local actor and critic. The critic part uses an exponential weighted average (EWA) for evaluating the actions of the actor part. Based on the evaluation of critic, the actor uses a linear parametric probability distribution for improving the routing policy for selecting the next-hop router. The simulation results show a reduced end-to-end delay and effectively balance the loads in a network. The existing literature for traffic management considers the routing optimization of the intra-domain routing. However, less work has been done to address optimization of inter-domain routing in an SDN network. The authors in \cite{zhao2020improving} proposed a multi-agent reinforcement learning (MARL) framework for addressing the issue of inter-domain routing and scalability. The consideration of inter-domain routing is useful for learning the optimal policy for collaborative networks running of large-scale distributed networks. The authors implemented the advantage actor-critic RL technique to optimize inter-domain routing with improving the system throughput and uses multiple inferences to reduce the size of action space. The Internet architecture is based on multiple autonomous systems (ASs), which interacts with each other for optimizing the traffic.  An AS in the proposed framework acts as an agent and collects the routing advertisements following the BGP policy. The training of the agent is done in a decentralized approach with its objective to maximize the system throughput. The proposed framework achieves good performance; however, for 5G beyond and envisioned 6G networks, accurate and efficient data-driven techniques are still needed to capture the non-linear relationship between inter-domain network.

\subsection{Traffic optimization based on RL and DT in 6G}

Future networks such as B5G and 6G will be based on highly complex and dynamic structures (such as IRSs and MIMO etc.,). Existing RL techniques for traffic optimization often lead to sub-optimal solutions as it is difficult to obtain accurate states from a dynamic and high-dimensional network. As a result, the DT layer combined with RL will create a virtual representation of the physical wireless network and accurate network states. They will be forwarded to the RL agent, which as a result will be able to learn the optimal traffic routing strategies for 6G networks\cite{keping}.

\subsection{Traffic optimization based on multi-task features}

The existing studies for TE are based on single features or tasks such as congestion control or load balancing and do not involve multiple task coordination. However, efficient traffic management techniques for envisioned 6G network may require multi-tasks. For example, the massive volume of traffic may require both load balancing and congestion control for traffic management. To address the issue, the authors in \cite{lei2020congestion} presented a multi-task scenario based on a deep reinforcement learning (DRL) approach. Compared to the single task technique, the congestion was taken as primary task and load balancing as the secondary task to learn the shared representation of the network. The problem of TE for multitask features is formulated as a continuous problem. A DDPG framework is proposed, where the multi-task DRL agent implemented in the control plane collects the networks' states and outputs an action based on load balancing and congestion control vectors. As a result, RL techniques based on multi-task features in the distributed control plane can optimize the uRLLC applications of 6G networks.

\subsection{Traffic optimization for solving the generalization issue}
Traffic optimization techniques based on a DRL framework cannot accurately approximate the action-value distribution due to the impact of random noise on the rewards. The generative adversarial network (GAN) has shown an improved performance on modeling the non-linear relationship directly from the data. The GAN is based on two models, i.e, generator and discriminator. The generator generates the plausible data and maps it to a real data. The discriminator receives the data from the generator and the real data and learns to distinguish the real data from the fake data. Both of the networks are trained using the gradient descent algorithm. The authors in \cite{gu2019ganslicing} presented a GAN approach for resource allocation in an SDN network. The GAN is implemented in the control plane that predicts the resource requirement of the nodes in a timely and dynamic way. A GAN based deep distributional Q-network (GAN-DDQN) framework is proposed in \cite{naeem2020generative} to address the scheduling issue in a highly dynamic and non-stationary environment. The deep distributional Q network based on a GAN is proposed to learn the action-value distribution. The generator network in the GAN generates a fixed number of particles and tries to match each output with the action-value distribution. The simulation result shows that in a non-stationary environment, the algorithm outperforms the existing DRL algorithms. Thus, RL algorithms enabled with GAN can solve the generalization issue in 6G networks.

\section{Future directions}
\begin{enumerate}

 \item  \textbf{Curse of dimensionality for large scale networks:}
  The DRL techniques used for TE performs well for networks having small state-action pairs. However, with the advent of envisioned 6G architectures, the networks will generate and forward a massive amount of traffic to the control plane with high dimensional state-actions pairs. As a result, searching for an optimal policy from high dimensions state-action space will degrade the performance of DRL in terms of converging. One of the future directions in the area of TE is to explore the concepts of pinning control, transfer learning, together with the DRL and DT to reduce the dimensions of large networks. By now, there is very limited research in the area of DRL with the application of pinning control to solve the problem of \textit{curse of dimensionality} in high dimensional networks. The future work can further explore the success of control theory implemented in the control plane and transfer learning with the DRL and DT approach for solving the scalability issues in envisioned 6G networks.

  \item  \textbf{Mobility management and forensic for IRS-assisted 6G network:}
  One of the key challenges in the DT network will be how to design the edge or fog-based twin with highly mobile nodes in 6G specially for the case of IRSs. Mobile nodes must be served without interruption and seamlessly by the DT-enabled 6G network during the service requirement of the users. To cope with this issue, one future research direction is how to effectively migrate the service to a new twin object. This will require precise and accurate prediction of user mobility to find the optimal location of new twin objects for providing service. The envisioned DT-enabled 6G networks will have a variety of entities involved (e.g., communication interfaces, end devices and TVM-based twin objects). As a result, the DT will be vulnerable to security attacks. Thus, effective forensic techniques such as blockchain are needed to mitigate the effect of such attacks.

  \item  \textbf{Efficient optimization techniques for digital twins:}
  The digital twin will construct a virtual representation of the physical wireless network where the network states such as IRSs, base-stations, and UAVs will be forwarded to the RL agent. As a result, one of the key challenges in DT will be to explore efficient online optimization and ML techniques that can dynamically update the data from an environment and keep the physical network synchronized with the DT network.

\item \textbf{Information centric networking for reducing redundant traffic:} 
The distributed nature of the ICN approach becomes very ineffective when handling global resource allocation and intensive computation tasks in a massive scale 6G networks. The content distribution in ICN involves caching networks across heterogeneous networks, which models a highly stochastic and dynamic process. The collection of data involves privacy and security issues and remains a future research direction in the context of SDN. Another future direction is to explore the data-driven techniques for deep learning considering the quantity and quality of the data. Thus, to achieve higher accuracy, domain specific models should be explored for the ICN in communication networks.
\item \textbf{Trial and error strategies for DRL:} 
   The TE problem for dynamic and complex structures such as (IRSs, UAV, MIMO etc.,) involves online-decision making. The existing DRL problems will not be able to adapt to the characteristics of the 6G communication systems. The DRL agent learns the optimal behaviour of a network by a trial-and-error method. The agent optimizes its behaviour by using a function approximation according to the penalties or rewards received when interacting with the environment. However, the envisioned 6G architecture will have uRLLC with stringent QoS requirements where the trial-and-error method is not an appropriate choice. Thus, stringent QoS requirements face an unprecedented research challenge for designing DRL models for the communication networks. One of the future directions will be to use efficient multi-agent exploration strategies in the envisioned 6G where agents learn the policies in a coordinated manner with shared goals of improving the performance of the network.

\item \textbf{Multi-agent reinforcement learning for envisioned 6G networks:}  
The multi-agent reinforcement learning (MARL) is a recent technique that involves set of agents and the objective, and optimal policy of the network. It is based on the joint actions of all the agents. One of the key challenges in the MARL is the non-stationary policies implemented in the heterogeneous networks. The future research direction is to design an approach that will collect the joint action space information of all agents in the training process. However, this will arise scalability issues due to a large number of state-action space. Techniques such as learn factorized value function with respect to actions can be explored in future to overcome the issues for MARL techniques using the SDN approach. The training in the MARL techniques is done in a centralized way while the execution of the policy using a distributed approach. Thus, the traffic orchestrator can be intelligently used for centralized training in MARL for optimizing the resources of the 6G networks.
\item \textbf{Security and privacy:}
The DRL techniques for traffic management based on single and multi-agent techniques may involve sharing information between different agents. For example, in case of IRSs, the traffic will be forwarded using multi-IRS elements. This will arise security and privacy concerns as sharing the rewards between different networks can lead to vulnerabilities and cyber-attacks. Thus a future direction will be to explore techniques like federating learning to overcome and mitigate different security and privacy challenges in the envisioned 6G networks.

\end{enumerate}

\section{Conclusion}

ML, especially the RL approach, has proved to be potentially the ultimate solution for achieving intelligence in the envisioned 6G networks. In this paper, we have proposed a futuristic digital twin empowered RL-aided traffic-orchestration architecture for optimizing the network performance of uRLLC, IoE, and  eMBB applications. We have also discussed the existing traffic orchestrator techniques empowered with the RL techniques used for optimizing the network performance. Finally, we have  highlighted several potential and promising research directions for the traffic management and orchestration in the envisioned 6G networks.


%

\ifCLASSOPTIONcaptionsoff
  \newpage
\fi



%

\bibliographystyle{IEEEtran}
\bibliography{IEEEabrv,Bibliography}

%




\end{document}